\documentclass[letterpaper, 10 pt, conference]{ieeeconf}  % Comment this line out if you need a4paper

\usepackage{siunitx}
\usepackage{amsmath, amsfonts, amssymb}
\usepackage{bm}
\usepackage{graphicx}

\IEEEoverridecommandlockouts                              % This command is only needed if 
\title{\LARGE \bf
Real-Time Optimization of the Current Steering for Visual Prosthesis*
}

\author{Zhijie Charles Chen$^{1}$, Bing-Yi Wang$^{2}$, and Daniel Palanker$^{3}$% <-this % stops a space
\thanks{*Research supported by the National Institutes of Health (Grants R01-EY-027786, P30-EY-026877), the Department of Defense (Grant W81XWH-19-1-0738), AFOSR (Grant FA9550-19-1-0402), Wu Tsai Institute of Neurosciences at Stanford.}% <-this % stops a space
\thanks{$^{1}$Z. C. Chen is with the Department of Electrical Engineering and Hansen Experimental Physics Laboratory at Stanford University, Stanford, CA 94305 USA 
        {\tt\small zcchen@stanford.edu}}%
\thanks{$^{2}$B.-Y. Wang is with the Department of Physics and Hansen Experimental Physics Laboratory at Stanford University, Stanford, CA 94305 USA
        {\tt\small bingyiw@stanford.edu}}%
\thanks{$^{3}$D. Palanker is with the Department of Ophthalmology and Hansen Experimental Physics Laboratory at Stanford University, Stanford, CA 94305 USA
        {\tt\small palanker@stanford.edu}}%
}

\begin{document}

\maketitle
\thispagestyle{empty}
\pagestyle{empty}

%%%%%%%%%%%%%%%%%%%%%%%%%%%%%%%%%%%%%%%%%%%%%%%%%%%%%%%%%%%%%%%%%%%%%%%%%%%%%%%%
\begin{abstract}

Current steering on a multi-electrode array is commonly used to shape the electric field in the neural tissue in order to improve selectivity and efficacy of stimulation. Previously, simulations of the electric field in tissue required separate computation for each set of the stimulation parameters. Not only is this approach to modeling time-consuming and very difficult with a large number of electrodes, it is incompatible with real-time optimization of the current steering for practical applications. We present a framework for efficient computation of the electric field in the neural tissue based on superposition of the fields from a pre-calculated basis. Such linear algebraic framework enables optimization of the current steering for any targeted electric field in real time. For applications to retinal prosthetics, we demonstrate how the stimulation depth can be optimized for each patient based on the retinal thickness and separation from the array, while maximizing the lateral confinement of the electric field essential for spatial resolution.

\end{abstract}

%%%%%%%%%%%%%%%%%%%%%%%%%%%%%%%%%%%%%%%%%%%%%%%%%%%%%%%%%%%%%%%%%%%%%%%%%%%%%%%%
\section{Introduction}

For restoration of vision to the blind, an array of electrodes is placed in either the epiretinal or subretinal location, or on the visual cortex of the patient. Prosthetic vision of high fidelity requires selective neural stimulation, which is limited by the density of the electrode array and the crosstalk between neighboring electrodes. Crosstalk can be minimized by placement of the local return electrodes in each pixel, but this reduces the penetration depth of electric field into tissue. As a result, retinal stimulation with pixels smaller than about \SI{50}{\micro\meter} becomes nearly impossible \cite{ho2019characteristics}. Crosstalk can also be reduced by time multiplexing – separating a video frame into multiple sparser sub-frames applied sequentially. However, the number of sub-frames is limited by the frame rate and by the pulse width required for neural stimulation. Another challenge is that neurons in the retina vastly outnumber the electrodes in the array, which limits the number of selective targets for stimulation. For example, the state-of-the-art photovoltaic retinal prosthesis (PRIMA, Pixium Vision) with \num{378} pixels targets an area of \SI{2}{\milli\meter}$\times$\SI{2}{\milli\meter} in the macula \cite{palanker2020photovoltaic}, populated by at least \num{40000} bipolar cells. Such discrepancy is even more significant for cortical implants, where the interface is 3-dimensional.

Current steering, a technique that utilizes simultaneous injection of current through multiple electrodes to shape the electric field in tissue, has been introduced to improve either the selectivity or efficacy of stimulation. It is used in cochlear implants \cite{bonham2008current}\cite{firszt2007current}\cite{wu2013current}, in deep-brain stimulation (DBS) \cite{choi2011modeling}\cite{chaturvedi2012current}\cite{barbe2014multiple}\cite{zhang2020comparing} and in suprachoroidal implants \cite{dumm2014virtual}\cite{spencer2018creating}\cite{matteucci2013current} to enhance the stimulation capabilities with very limited number of electrodes, including creation of the “virtual electrodes” - centers of stimulation between the physical electrodes \cite{firszt2007current}\cite{dumm2014virtual}\cite{spencer2018creating}.  It was also used to direct current across the highly resistive retinal pigmented epithelium (RPE) layer \cite{matteucci2013current}. In cortical visual prosthesis, current steering was implemented to provide smooth transition between the sequentially activated electrodes \cite{beauchamp2020dynamic}.

Computation of the electric field emanating from multiple electrodes involves Poisson’s equation of volume conduction, which can be solved with the finite element method (FEM) \cite{bonham2008current}\cite{choi2011modeling}\cite{chaturvedi2012current}\cite{dumm2014virtual}\cite{spencer2018creating}\cite{matteucci2013current}. In principle, for a certain optimization of the field confinement, the FEM, which takes at least several seconds to run on a typical computer, needs to be repeated for a very large number of the current settings on multiple electrodes. The computational cost increases with the total number of possible configurations, and would be intractable with even a few electrodes, each having tens of possible levels of current injection. Although iterative algorithms, such as gradient descent, may accelerate the convergence to the optimum, the FEM still needs to run for a number of times, thus precluding the FEM-based optimization for a video stream in real time, where the frame rate is typically in tens of \si{\hertz} and the computation must be completed within tens of milliseconds.

Since electric conduction in the tissue is typically linear (conductivity does not change with electric field) \cite{chen2020harmonic}, and by the principle of superposition, the electric field is a weighted sum of the electric fields generated by each electrode individually, we may construct a dictionary comprising the elementary electric fields produced by each electrode with a unitary current, and calculate the actual electric field by linear combination of all dictionary entries, with coefficients equal to the respective currents. Similar approach was taken by \cite{dmochowski2011optimized} in modeling transcranial direct current stimulation. Although initial computation is required to generate the dictionary, all possible electric fields could be rapidly synthesized without further running the FEM. Given the specifications of a targeted electric field, we can then find the best approximation of the current steering scheme under the minimum-mean-square-error (MMSE) criterion. We demonstrate a utility of such approach for subretinal implants to find the optimal activation schemes that extend the stimulation into different depths of the inner nuclear layer (INL), while maintaining lateral confinement of electric field for high spatial resolution.

\section{The System Model}
\subsection{A Basis for the Electric Field Computations}

Let us consider an array of $M$ active electrodes interfacing with a volume of neural tissue $T$, which is the retina or the visual cortex, depending on the application. The surfaces with the electrodes in $T$ are denoted by $S_1, S_2, \cdots, S_M$ with surface areas of $s_1, s_2, \cdots, s_M$, respectively. $\varphi(\bm{r})$ denotes the potential distribution in the tissue as a function of the spatial variable $\bm{r}$. We choose $\varphi(\|\bm{r}\|\to+\infty)=0$ and define $\hat{\bm{n}}_m$ as the normal to $S_m$, where $m\in\{1, 2, \cdots, M\}$.

The current injection of the $m^{\text{th}}$ electrode $I_m$ sinks to the return electrode $S'_m$, with area $s'_m$ and normal $\hat{\bm{n}}'_m$. All the return electrodes may be connected as a common ground, so the notation does not imply that pixels are organized into bipolar pairs of the active and return electrodes. We assume a uniform current density (secondary current distribution) at the electrode-electrolyte interfaces, which is common in neural stimulation \cite{chen2020current}. The framework can be easily adapted for other boundary conditions at the electrode surface, such as the primary or tertiary current distribution \cite{wang2019review}. We also adopt the convention of $I_m>0$ for anodal current.

Let $\bm{\sigma}$ be the conductivity of the neural tissue, which may vary spatially in a non-uniform tissue, or be a tensor in anisotropic medium \cite{zhang2020comparing}. By the Poisson’s equation for volume conduction, we have
\begin{equation}\label{eq_poisson}
    \nabla(\bm{\sigma}\nabla\varphi)=0.
\end{equation}
The boundary conditions are
\begin{subequations}\label{eq_boundary}
\begin{equation}\label{eq_inf}
\lim_{\|\bm{r}\|\to+\infty}\varphi(\bm{r})=0,
\end{equation}
\begin{equation}
-(\bm{\sigma}\nabla\varphi)\cdot\hat{\bm{n}}_m=\dfrac{I_m}{s_m} \quad \text{at } S_m,
\end{equation}
and
\begin{equation}
-(\bm{\sigma}\nabla\varphi)\cdot\hat{\bm{n}}'_m=-\dfrac{I_m}{s'_m} \quad \text{at } S'_m
\end{equation}
for all $m$, and
\begin{equation}
    (\bm{\sigma}\nabla\varphi)\cdot\hat{\bm{n}}=0
\end{equation}
\end{subequations}
at all inactive surfaces. In practice, (\ref{eq_inf}) can be imposed by a grounded bounding box that is much larger than the region of interest (ROI).

The equation system defined by (\ref{eq_poisson}) and (\ref{eq_boundary}) is linear with respect to $I_m$. Therefore, all possible electric fields in the neural tissue constitute an $M$-dimensional vector space. Let $I_0$ be the unitary current, and $\varphi_k$ be the potential distribution defined by (\ref{eq_poisson}) and (\ref{eq_boundary}) when
\begin{equation}
    I_m=\begin{cases}
  I_0,\quad \text{if } m=k;\\    
  0,\quad \text{  if } m\neq k,    
\end{cases}
\end{equation}
which is the potential distribution generated by the $k^\text{th}$ electrode injecting $I_0$ of current individually. $\{\varphi_1, \varphi_2, \cdots,\varphi_M\}$ forms a basis of the vector space. By the principle of superposition, we have
\begin{equation}
    \varphi=\frac{1}{I_0}\sum_m\varphi_mI_m.
\end{equation}

Thereby, we construct a dictionary of the elementary electric fields $\{\varphi_m\}$, from which all possible electric fields can be synthesized by linear combination with the currents at the electrodes as the coefficients. Note that the elementary electric fields are specific to the application, which may not require the granularity of a single electrode. For example, when the \SI{40}{\micro\meter}-pixel photovoltaic subretinal implant described by \cite{huang2020vertical} is used to determine grating acuity, it involves line-by-line activation of pixels. In this case, the dictionary consists of the electric fields generated by all the pixels uniformly activated in each line. This reduces the number of entries in such a dictionary from \num{425} (the number of pixels) to only \num{35} (the number of lines).

Storage of the dictionary can be resource-demanding, especially with a large number of electrodes. However, rarely is the case that potential distribution in the entire space is of interest. Usually only a subset of metrics is relevant for the application, such as the potential drop across a certain region. As long as the metrics are linearly derived from the electric field, storage of the entire electric field is unnecessary. We can first derive the metrics from the elementary electric fields, and then construct the dictionary with only the metrics of interest. For example, if we are interested in the potential difference $\Delta V$ between the dendritic and the axonal ends of a bipolar cell, the dictionary only needs to store the potential difference $\Delta V_m$ derived from each elementary electric field $\varphi_m$, and $\Delta V$ is given by
\begin{equation}
    \Delta V = \frac{1}{I_0}\sum_m\Delta V_mI_m.
\end{equation}

\subsection{Model of a Subretinal Implant} \label{subsect_2B}

\begin{figure}[thpb]
      \centering
      \includegraphics[width=\columnwidth]{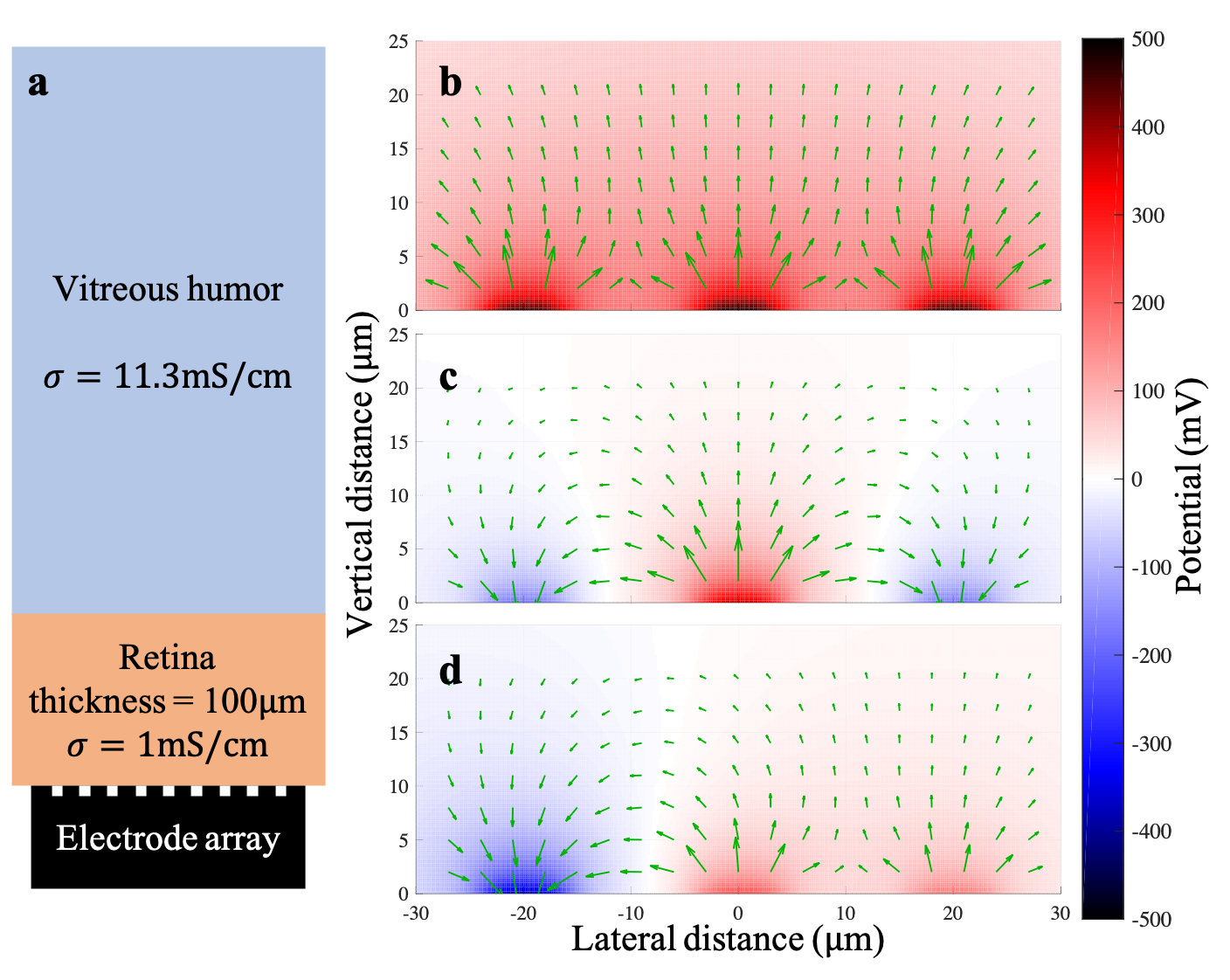}
      \caption{(a) Schematic diagram of a subretinal implant in the eye. (b) Monopolar stimulation with \SI{0.5}{\micro\ampere} of current at each electrode. (c) Bipolar stimulation with \SI{0.5}{\micro\ampere} at the center electrode, and \SI{-0.25}{\micro\ampere} at each of the side electrodes. (d) Virtual electrode created at the center-right by \num{-0.5}, \num{0.3} and \SI{0.2}{\micro\ampere} at the left, center and right electrode, respectively. (b)-(d) are synthesized by linear combination of the three elementary electric fields.}
      \label{fig_1}
   \end{figure}

Fig. \ref{fig_1}a illustrates a subretinal implant with active electrodes arranged in a 2-dimensional grid of \SI{20}{\micro\meter} in pitch, with a common distal ground electrode. Each active electrode is a disk of \SI{8}{\micro\meter} in diameter, and the electrode array is \SI{1.5}{\milli\meter}$\times$\SI{1.5}{\milli\meter} in size, resembling the \SI{20}{\micro\meter}-pixel device described in \cite{huang2020vertical}. The retina has a thickness of \SI{100}{\micro\meter}, a conductivity of \SI{1}{\milli\siemens\per\centi\meter} \cite{werginz2020optimal} and conductivity of the vitreous body above the retina is \SI{11.3}{\milli\siemens\per\centi\meter} \cite{oksala1959comparative}.

The stimulation paradigm is configured by the currents at several neighboring electrodes. For example, let us consider a central electrode and the two neighboring electrodes on both sides. The dictionary will contain three elementary fields, from which we can synthesize various linear combinations for the stimulation paradigms.

   When current of the same polarity is passed through each electrode, the stimulation is monopolar, with deep penetration and minimal lateral confinement of the electric field (Fig. \ref{fig_1}b). When currents at the side electrodes have half of the amplitude and opposite polarity of the central electrode, the stimulation is bipolar (Fig. \ref{fig_1}c), with the electric field penetrating shallower into the retina but more confined laterally. Changing the ratio of currents between the side electrodes shifts the center of stimulation away from the central electrode, and hence creating the putative virtual electrode (Fig. \ref{fig_1}d). Adjusting the proportion of current sinking to the side electrodes versus to the distal ground yields intermediate paradigms between monopolar and bipolar, and thus enables laterally confined electric fields of various penetration depths, which we will discuss in Section \ref{subsect_3C}.

\section{Real-Time Optimization}

To optimize the current steering scheme for each frame of the video stream, we need to specify the desired electric field. For example, a subretinal implant should depolarize the bipolar cells in the target region by creating a potential difference between the dendritic and the axon-terminal ends of the cells \cite{werginz2020optimal}, while keeping the depolarization outside the target region as low as possible. Each scalar metric of the electric field is one specification (e.g. the potential at a certain point or the potential difference between two points), but a 3-dimensional variable (e.g. the current density at a point) should be considered three specifications. Depending on the number of specifications, the optimization falls in two regimes: underdetermined or overdetermined.

\subsection{The Underdetermined Problem} \label{subsect_3A}

When the number of specifications is smaller than the number of independent electrodes, the problem is underdetermined. Since we have more “tuning knobs” than specifications, all specifications can be satisfied and we may consider other optimization criteria, such as minimizing the energy consumption, for example.

For illustration, let us consider the problem of defining the potential differences between the dendritic and the axon-terminal ends of bipolar cells at $L$ different locations, with $M$ subretinal electrodes, where $L<M$. Let $V_l$ be the specification for the potential difference at the $l^\text{th}$ location, and $u_{l,m}$ be the potential difference at the $l^\text{th}$ location in the elementary electric field $\varphi_m$. Define $\bm{U}$ as the matrix with entry $u_{l,m}$ at row $l$ and column $m$, and $\bm{v}$ as the column vector whose $l^\text{th}$ component is $V_l$. We aim to find the currents at the electrodes $\bm{x}={\left[I_1, I_2, \cdots, I_M\right]}^\intercal$ such that
\begin{equation} \label{eq_ud}
    \bm{U}\bm{x}=\bm{v}.
\end{equation}

The solution to (\ref{eq_ud}) is given by
\begin{equation} \label{eq_ud_sol}
    \bm{x}^*=\bm{U}^+\bm{v}+\bm{w},
\end{equation}
where $\bm{U}^+=\bm{U}^\intercal{\left(\bm{U}\bm{U}^\intercal\right)}^{-1}$ is the pseudoinverse of $\bm{U}$ and $\bm{w}$ is any vector in the nullspace of $\bm{U}$. All specifications are satisfied, and the solution is not unique, allowing us to optimize for other criteria. For example, the sum of squares of all currents reflects the total power consumption. Therefore, minimal power consumption is achieved if and only if $\bm{w}=\bm{0}$.

\subsection{The Overdetermined Problem} \label{subsect_3B}

When there are more specifications than electrodes, the problem is overdetermined. Generally, all specifications cannot be met simultaneously, and we aim to find the electric field closest to the specifications under the MMSE criterion, by finding the optimal current steering scheme $\bm{x}^*$, such that ${\|\bm{U}\bm{x}-\bm{v}\|}^2$ achieves the minimum. The optimal scheme is given by
\begin{equation} \label{eq_od_sol}
    \bm{x}^*=\bm{U}^+\bm{v},
\end{equation}
where $\bm{U}^+={\left(\bm{U}^\intercal\bm{U}\right)}^{-1}\bm{U}^\intercal$ is the pseudoinverse of $\bm{U}$.

In both cases, the pseudo-inversion $\bm{U}^+$ is pre-calculated and does not change with the image frame. The real-time computation for the optimal current steering scheme involves only one matrix-vector multiplication, which makes the computation highly efficient.

\subsection{Stimulation Depth and Lateral Field Confinement} \label{subsect_3C}

To achieve prosthetic visual acuity matching the sampling limit of the electrode array, lateral confinement of the electric field is required to suppress crosstalk between the electrodes. However, lateral confinement with local returns also limits the penetration of electric field into the retina. Small bipolar pixels over-constrain the field penetration to less than a pixel radius and thus strongly diminish the stimulation efficacy \cite{flores2019honeycomb}. We may laterally confine the electric field more efficiently with current steering. Using the theory developed in Section \ref{subsect_3B}, we will optimize for the current steering schemes for selective stimulation of the INL at various depths with the subretinal implant described in Section \ref{subsect_2B}.

Let us assume that activation of a bipolar cell is determined by the potential difference between its two ends \cite{werginz2020optimal}, and that its length is \SI{35}{\micro\meter}. Let us also assume that proximity between the implant and the dendritic ends of the bipolar cells, denoted by $z_0$, varies between patients, and we will model three scenarios where $z_0$ equals \num{5}, \num{25} and \SI{45}{\micro\meter}, respectively. The ROI is a \SI{100}{\micro\meter}-wide square concentric to the implant, in which we configure the potential difference  between $z_0$ and $z_0+$\SI{35}{\micro\meter} with \num{141} electrodes near the center of the implant. The targeted potential difference is a 2-dimensional boxcar function, where the value is \SI{50}{\milli\volt} in the central square of \SI{40}{\micro\meter} in width and \SI{0}{\milli\volt} elsewhere. The ROI is discretized in steps of \SI{0.5}{\micro\meter}, yielding \num{201}$\times$\num{201} sampling points and thus \num{40401} specifications. The problem is overdetermined, and we aim to find the optimal current steering schemes that selectively activate the central square at different depths.

\begin{figure}[thpb]
      \centering
      \includegraphics[width=\columnwidth]{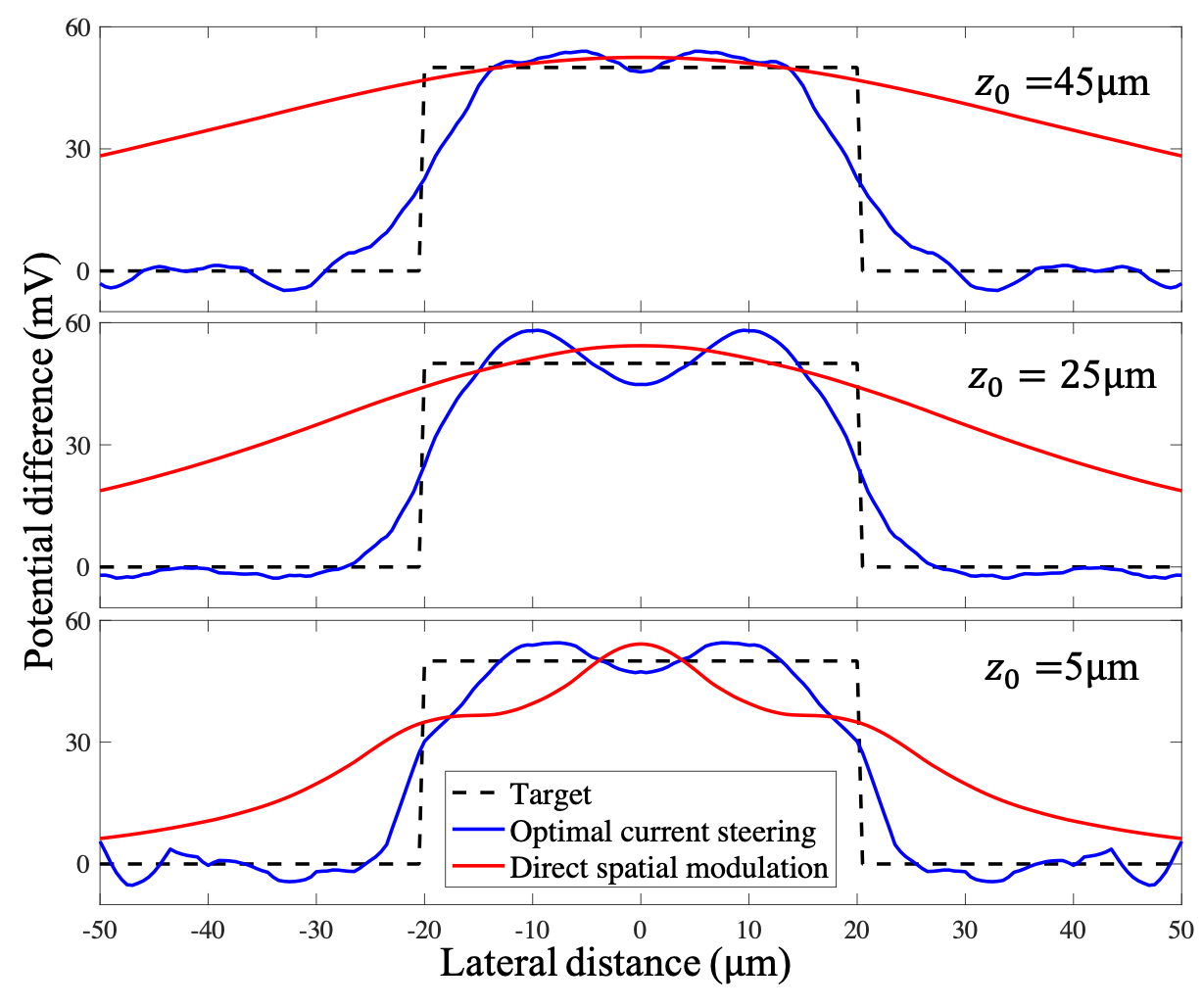}
      \caption{Lateral confinement of the electric field at three stimulation depths. Top: potential difference between \num{45} and \SI{80}{\micro\meter} from the implant; Middle: \num{25} to \SI{60}{\micro\meter}; Bottom: \num{5} to \SI{40}{\micro\meter}.}
      \label{fig_2}
   \end{figure}

Fig. \ref{fig_2} compares the lateral confinement of the electric field by the optimal current steering at three stimulation depths, with that by direct spatial modulation, where only the electrodes in the central square are activated. Optimal current steering confines the electric field much better at all stimulation depths, especially at deeper target layers.

\section{Discussion}

Equations (\ref{eq_ud_sol}) and (\ref{eq_od_sol}) show that the real-time optimization for each frame requires only one matrix-vector multiplication, which is a basic operation that has been extensively optimized in many software packages. The optimization in Section \ref{subsect_3C} involves \num{141} electrodes and \num{40401} specifications of the targeted electric field, yet takes only \SI{2}{\milli\second} to run on a laptop computer (\SI{3.1}{\giga\hertz} dual-core CPU, \SI{8}{\giga B} memory) with MATLAB (\num{64}-bit R2017a, Mathworks Inc.), meeting the requirement for real-time implementation. The matrix-vector multiplication can further benefit from hardware acceleration using field-programmable gate array (FPGA) or application-specific integrated circuit (ASIC).

Although optimization of the current steering we propose is highly efficient, construction of the dictionary could be time consuming with a large number of electrodes, since each elementary electric field still requires one FEM simulation. However, we may exploit the symmetry and invariance in the geometry of the electrode array to simplify the process. For example, for the optimization in Section \ref{subsect_3C}, since all the engaged electrodes are near the center of an implant which is much larger than the ROI, the edge effects are negligible, and thus the elementary fields are space-shifted versions of each other. Thereby, the FEM simulation only needs to run once.

In patients with subretinal implants, the critical trade-off between the stimulation depth and lateral confinement of the electric field depends on the proximity between the INL and the implant, which varies between patients and is affected by multiple factors, such as post-surgical retinal reattachment, potential retinal edema, state of degeneration, and others. With local returns around each active electrode to confine the field, different patients would require implants with different pixel sizes that may not even be known before the implantation. The optimal current steering provides an approach to adjustment of the stimulation depth by software, thus accommodating all patients with a single design of the implant.

With the optimal current steering, deep penetration of the electric field does not necessarily compromise the lateral selectivity. Section \ref{subsect_3C} demonstrates a boxcar function of the same width at different depths of the INL. A sharp boxcar function is an important building block for representing a complex image as multiple sparse sub-frames, which are sequentially activated within a frame duration. The maximum number of sub-frames is determined by the ratio of the frame duration ($1/\text{frame rate}$) to pulse duration required for neural stimulation. The optimization we propose enables sufficiently fast computation to fit the requirement not only for a video rate, but also for the time multiplexing with the sub-frame duration in the range of a few milliseconds.

This framework and the optimization may also be used in other areas of neural stimulation, such as cochlear implants and DBS, for example.

\section{Conclusion}

A framework was proposed to efficiently model and optimize for the current steering in neural tissues by linear combination of a set of pre-calculated elementary electric fields. Such approach to finding the optimal current steering scheme for any targeted electric field can be implemented in real time. We demonstrated a utility of such optimization for subretinal implants to stimulate the INL at different depths while maintaining a constant lateral confinement, thereby overcoming the limitation on penetration depth of the electric field imposed by the fixed local return electrodes.

\addtolength{\textheight}{-13cm}   % This command serves to balance the column lengths

%%%%%%%%%%%%%%%%%%%%%%%%%%%%%%%%%%%%%%%%%%%%%%%%%%%%%%%%%%%%%%%%%%%%%%%%%%%%%%%%

\bibliographystyle{IEEEtran}
\bibliography{IEEEabrv, ref}

\end{document}